\documentclass[preprint,12pt,authoryear]{elsarticle}

\usepackage{graphicx,amssymb,amsfonts,amsmath,color,rotating,multirow,array,setspace}
\usepackage[percent]{overpic}
\makeatletter
\def\ps@pprintTitle{%
 \let\@oddhead\@empty
 \let\@evenhead\@empty
 \def\@oddfoot{\centerline{\thepage}}%
 \let\@evenfoot\@oddfoot}
\makeatother

\makeatletter
\g@addto@macro{\endtabular}{\rowfont{}}
\makeatother
\newcommand{\rowfonttype}{}
\newcommand{\rowfont}[1]{
   \gdef\rowfonttype{#1}#1%
}
\newcolumntype{L}{>{\rowfonttype}l}

\begin{document}
\begin{frontmatter}



\title{Interdependencies of female board member appointments}

\author[label1,label2,label3]{Matthias Raddant}
\author[label4]{Hiroshi Takahashi}
\address[label1]{Kiel University, Department of Economics, 24118 Kiel, Germany}
\address[label2]{Danube University Krems, Department for Knowledge and Communication Management,
3500 Krems, Austria}
\address[label3]{Complexity Science Hub Vienna, 1080 Vienna, Austria}
\address[label4]{Keio University, Graduate School of Business Administration,\\ Yokohama Kanagawa, 223-8526, Japan
}

\begin{abstract}

We investigate the networks of Japanese corporate boards and its influence on the appointments of female board members. We find that corporate boards with women show homophily with respect to gender. The corresponding firms often have above average profitability. We also find that new appointments of women are more likely at boards which observe female board members at other firms to which they are tied by either ownership relations or corporate board interlocks.
 
\end{abstract}


\end{frontmatter}
\section{Introduction}

The continuing under-representation of women on corporate boards is a complex problem. Studies find that gender-related biases often start in the education system and continue during the career paths of young adults and early professionals \citep{bram,jadidi,page}, often leading to less and less women in the upper hierarchies of certain professions. In some cases, especially for high-level appointments like corporate board members, quotas may help to improve the gender balance. However, we argue that if we want to achieve lasting changes, we have to better understand the dynamics in the appointments of female board members and the interdependencies between the appointment decisions of different boards. We also have to investigate if any relationship exists with the firm's financial performance.  Hence, this study focuses on the last steps in the career paths of female board members. It is based on the small group of women on the boards of 4,000 Japanese corporates. We analyze which companies appoint them, and we investigate what determines the slow but steady growth in the number of female board members from 2004 until 2013. In particular we argue that the number of female board member appointments depends to a significant degree on interaction effects in the practices of connected boards.

The share of female board members in Japan lies around 2 percent with a slight increase throughout the 10-year period covered by our data set. This is far below the numbers seen in other western economies \citep[see also][]{faccio}. Even in 2017 the share of female executives is far from the goal of 10 percent (or one per board) as announced by the Japanese government in 2013 \citep{genderE}. Hence, Japan does not have a quota for female board members but has announced clear policy goals about the increase of female labor market participation and representation in leading positions \citep[see also][]{saito}. 

The composition of corporate boards in general has been a subject of research for quite some time. To much surprise, it seems that the composition is often rather irrelevant for the success of a company, provided that basic governance practices are met \citep{dalton}. However, only very few studies actually investigate the gender ratio on corporate boards \citep[see also][]{femita,kirsch18}. 

The findings on the influence of gender on performance differ.
For example, \cite{gue} finds that female CEOs under-perform with respect to shareholder's returns. \cite{ahdit} find that the increase in female board members might have led to less experienced boards and thus decreased performance.  \cite{camp} however find the opposite effect. From an investors's point of view the gender does not seem to play a role either \citep{brink}. With respect to earnings volatility and firm performance \cite{faccio} and \cite{cony17} find positive effects from board diversity and female CEOs.

Studies like \cite{galbreath} argue that the link between the representation of women on boards of directors and financial performance is an indirect one, and that board room gender is associated with measures of corporate social responsibility which in turn positively influences financial performance. Several other studies take related viewpoints and argue
 that while the under-representation of woman on boards might not be much of an issue in the short run, there is evidence that it influences the pace of innovation within the company \citep{femnorw,fem_rep}, as well as the setting of societal norms and practices within the corporation and outside of it \citep[see also][]{matsa,kirsch21,cimpian,dezso}.\footnote{Similar observations have also been made about diversity in the board room in general, see for example \cite{board_diverse}.}
Hence, the appointments of female board members serve as an indicator for gender equality and social responsibility and are likely to influence the career paths of future generations of female executives \citep[see also][]{bilimoria}. 

The ongoing process of new appointments takes place in a networked environment where managerial practices, governance and appointment decisions are influenced by what is observed in other corporations \citep{Borgatti03,uzzi,branson}.
The interorganizational networks within which these managerial practices diffuse are manyfold: they stem from ownership relationships, supply chains, contacts to customers and ties between the board rooms of corporations \citep{gulati,Provan07}. Especially the case of corporate board interlocks has been analyzed in great detail. While the relationship between such interlocks and firm performance is debated, there is ample evidence in the literature about the diffusion of practices through such interlocks \citep{horton,zona,devos,lamb,mizruchi}.

For the case of Japan, firm networks have been analyzed from different perspectives. Firm networks based on production processes have for example been analyzed by \cite{son1}, firm conglomerates have been analyzed by \cite{dyad} and the interplay of board and ownership networks was analyzed by \cite{radhiro1}. While corporate governance in Japan in general was for example investigated by \cite{jpn_gov}, this study is to our knowledge, besides \cite{kubo}, the first large-scale analysis of female board members in Japan and one of the largest samples ever analyzed for western economies.

In the following we will analyze the evolution of the group of female board members and their networks. Our network-approach is partly motivated by studies that have shown that women often tend to connect to other women more than men to men, presumably due to in-group support \citep[see also][]{ingroup}. Additionally, the group of female executives of course forms a minority in an otherwise male-dominated network, which makes homophily one possible factor for the appointment of female board members \citep[see][]{karimirank,mcd}. 

We will therefore first analyze the networks of corporate boards with respect to differences between male and female board members. We will then test for the influence of the network on the presence and appointments of female board members.
In particular, we give an overview about the data and the distribution of mandates and ties in the board network in section \ref{sec:stats}. In section \ref{sec:nets} we discuss the extent of homophily in the different networks and the centrality of female board members compared to male ones. We then analyze the determinants of the number of female board members and the relationship with firm performance in sections \ref{sec:numfem} and \ref{sec:perf}. Section \ref{sec:hire} analyzes female board member appointments and how they are related to network effects.

\section{Board members and their networks}\label{sec:stats}
\subsection{The data}\label{sec:data}

For our analysis we have collected data of all publicly listed companies in Japan. Most of these are listed at the Tokyo Stock Exchange (TSE). This means that our sample includes all the roughly 1,700 firms of the so-called first section together with a similar amount of slightly smaller firms. We combine the data on the composition of corporate boards available from Toyo Keizai with financial data obtained from Nikkei Needs and Thompson Reuters Datastream. In particular we use the information on market value, income, total assets, the business sector, largest shareholders and share holder composition.

The information on the composition of the board is updated annually in the middle of the year. Besides the names of the board members we know their age, gender and whether they are outside board members or auditors. The naming and numerical identifiers of board members are unanimous within each year, but not necessarily throughout the years. Therefore we have developed an algorithm that traces the destinies of board members over time based on parts of their names, date of birth and affiliations.\footnote{We have confirmed the validity of this algorithm by manual checks. The only known limitation of this method is that we may lose traces of board members who exit the data set and re-appear at a later year at a different company. Manual checks have confirmed that for the following analysis this effect is negligible.} The financial data of the firms is matched using the same yearly frequency \citep[see][for a more general analysis of this data set]{radhiro1}.

In Japan boards are often comprised of about 10 corporate executive officers plus 2 to 3 externals. The exact size of the board varies and is related to firm size. Hence, typically we observe cases where the board of directors consists of 6 to 15 corporate executive officers, 1--2 auditors and possibly 1 or 2 outside board members. Only few mainly very large corporations report up to 35 total board members.

Networks between board members and firms can be constructed in the following way: for each year we observe a set of board members and a set of firms. Board members serve on the boards of one or more firms. This creates relationships (incidences) between the set of board members and the set of firms which resemble a bi-partite graph. Incidences can be described by positive entries in the incidence matrix $I$, where the dimensions of $I$ are given by the number of firms and the number of board members.

From the incidence matrix $I$ we can obtain two different un-directed networks by projection. 
$A_P = I I'$ creates an adjacency matrix for the \emph{personal network} of board members, where positive entries resemble cases where board members know each other from serving on at least one board together. By multiplying  $I' I = A_B$ we obtain an adjacency matrix that describes the network of the firms based on board interlocks. In the following we will refer to this network as the \emph{board network}.

\begin{sidewaystable}[p]
\footnotesize
\setlength{\tabcolsep}{4pt}
  \begin{center}
    \begin{tabular}{ c c c c c c c c c c c}
      \hline
      \hline

  & 2004 & 2005 & 2006 & 2007 & 2008 & 2009 & 2010 & 2011 & 2012& 2013  \\   
\hline
  no. firms & 3,767 & 3,849 & 3,943 & 3,887 & 3,767 & 3,672 & 3,595 & 3,543 & 3,532 & 3,545 \\ 
  no. b. members & 42,175 & 42,635 & 43,121 & 41,998 & 39907 & 38,759 & 37,731 & 36,884  & 36,452 & 36,697 \\ 
  no. mandates & 45,119 & 45,760 & 46,257 & 44,991 & 42,693 & 41,461 & 40,318 &  39,458 & 39,103 & 39,503 \\ 
 avg. board size & 11.98 & 11.89 & 11.73 & 11.57 & 11.33 & 11.29 & 11.22 & 11.14 & 11.07 & 11.14 \\
 \\
 avg. mandates & 1.070 & 1.073 & 1.073 & 1.071 & 1.070 & 1.070 & 1.069 & 1.070 & 1.073 & 1.076 \\
 avg. mand. women & 1.046 & 1.053 & 1.053 &	1.062 &	1.069 &	1.072 &	1.084 &	1.095 &	1.103 &	1.136 \\
 \\
 max mandates & 11 & 9 & 8 & 8 & 8 & 8 &7&  7& 7& 7 \\ 
 max mand women & 3 & 4 & 4 & 4 & 5 & 5 &5&  5& 5& 5 \\ 
\\
women & 439 & 506 & 533 & 515 & 493 & 514 & 536 & 571 & 623 & 713 \\  
\% women & 1.04 & 1.19 & 1.24 & 1.23 & 1.24 & 1.33 & 1.42 & 1.55 & 1.71 & 1.94 \\ 
\\ 
avg. age & 57.2 & 57.3 & 57.2 & 57.4 & 57.7 & 58.1 & 58.3 & 58.4 & 58.7 & 58.9 \\  
avg. age women & 51.3 & 50.9 & 51.6 & 52.0 & 52.3 & 52.8 & 52.9 & 53.3 & 53.8 & 54.1 \\ 
\\
\% auditors (men) & 17.18 & 18.00 & 18.42 & 18.89 & 19.27 & 19.64 & 19.74 & 19.86 & 19.84 & 19.63 \\
\% auditors (women) & 19.82 & 21.74 & 25.33 & 28.16 & 29.82 & 29.57 & 30.60 & 31.35 & 31.14 & 30.01 \\
\\
edges board net. & 3,221 & 3,498 & 3,415 & 3,266 & 3,045 & 2,956 & 2,883 & 2,869 & 2,984 & 3,240 \\  
edges owners. net.($>2$\%) & 1,920 & 2,047 & 2,162 & 2,138 & 2,167 & 2,417 & 2,451 & 2,528 & 2,604 & 2,667 \\
edges personal net. & 561,884 & 559,258 & 558,296 & 535,606 & 495,746 & 477,910 & 459,452 & 444,194 & 437,704 & 444,716 \\

\hline		
    \end{tabular}
  \end{center}
  \caption{Statistics on board members, firms, and networks connecting them}{\itshape While the number of firms shows some churning over time, the size of the average board is rather stable. The number of female board members is increasing significantly, their share however remains low. The increase in the average number of mandates for female board members hints at an increase of their importance.}\label{tab:bstats}
\end{sidewaystable}

We have also obtained data that reports the five largest shareholders (and their exact shareholding) for all of the firms in our sample on a yearly basis. This might at first seem a little restrictive, yet in practice significant influence onto a company is unlikely to be performed by more than five owners. Also, since this data is reported from the point of the owned company this still results in a rather complete picture of the ownership network.

In the following we will refer to the firm networks based on ownership simply as the \emph{ownership networks}. They differs from the board networks by the fact that they are directed networks. The densities are however comparable, the ownership network in 2004 contains 2,574 directed links and we see a steady increase until 2013 when the network has 3,695 links. 

Some statistics on these networks are given in table \ref{tab:bstats}. The number of female board members almost doubles with the observed period. This goes at hand with some other changes. The average number of mandates for female board members is catching up and is finally exceeding the overall average. Female board members are also over-represented in the group of auditors, this finding intensifies over time.

\subsection{Distributions of mandates and degree}\label{sec:mand}

The change in the number of mandates of female board members deserves further attention. In the following we will explore how frequent multiple mandates occur among female board members and how the distributions compare to those of male board members. We will also investigate how this translates to the degree distributions in the board network and thus to the connectivity of female board members.

\begin{figure}[htb]
\footnotesize
\begin{center}
 \includegraphics[width=0.75\textwidth,trim= 70 255 170 250, clip=true]{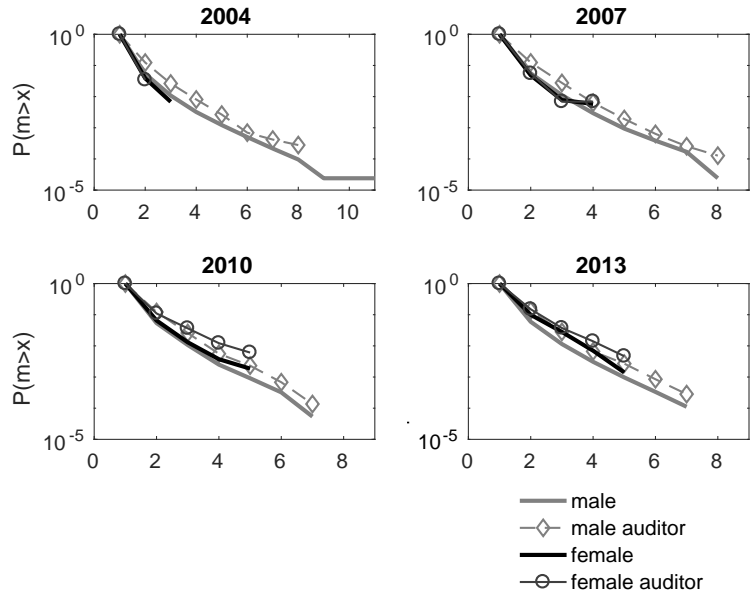}
 \end{center}
  \caption{Distributions of mandates}{\itshape We show the frequency of the number of mandates on a semi-log scale for male and female board members. Additionally we show these figure only taking auditors into account. We observe that the maximum of mandates for female board members increases over time and that their relative frequency surpasses those of men.}\label{fig:mand}
\end{figure}

Figure \ref{fig:mand} gives an overview about the development with respect to multiple mandates. We show the distributions for the number of mandates in 3-year intervals. Since female board members appear slightly more often in the role of the auditor than their male counterparts, we separately report the distributions that only consider auditors. In general, auditors are slightly more likely to have multiple mandates than the entire population of board members. For female auditors this only holds in the later years of our data set.
When we compare men and women we can clearly see a catching-up of the female board members in terms of the maximum  and the relative frequency of mandates. In 2004 there is no woman with more than 3 mandates, while in 2013 we observe a maximum of 5 (compared to 7 for men). More importantly, in 2013 women are more likely to have multiple mandates than men (which of course mirrors our observation from table \ref{tab:bstats}).

\begin{figure}[tb]
\footnotesize
\begin{center}
    \includegraphics[width=0.75\textwidth, trim = 30 370 180 140, clip=true]{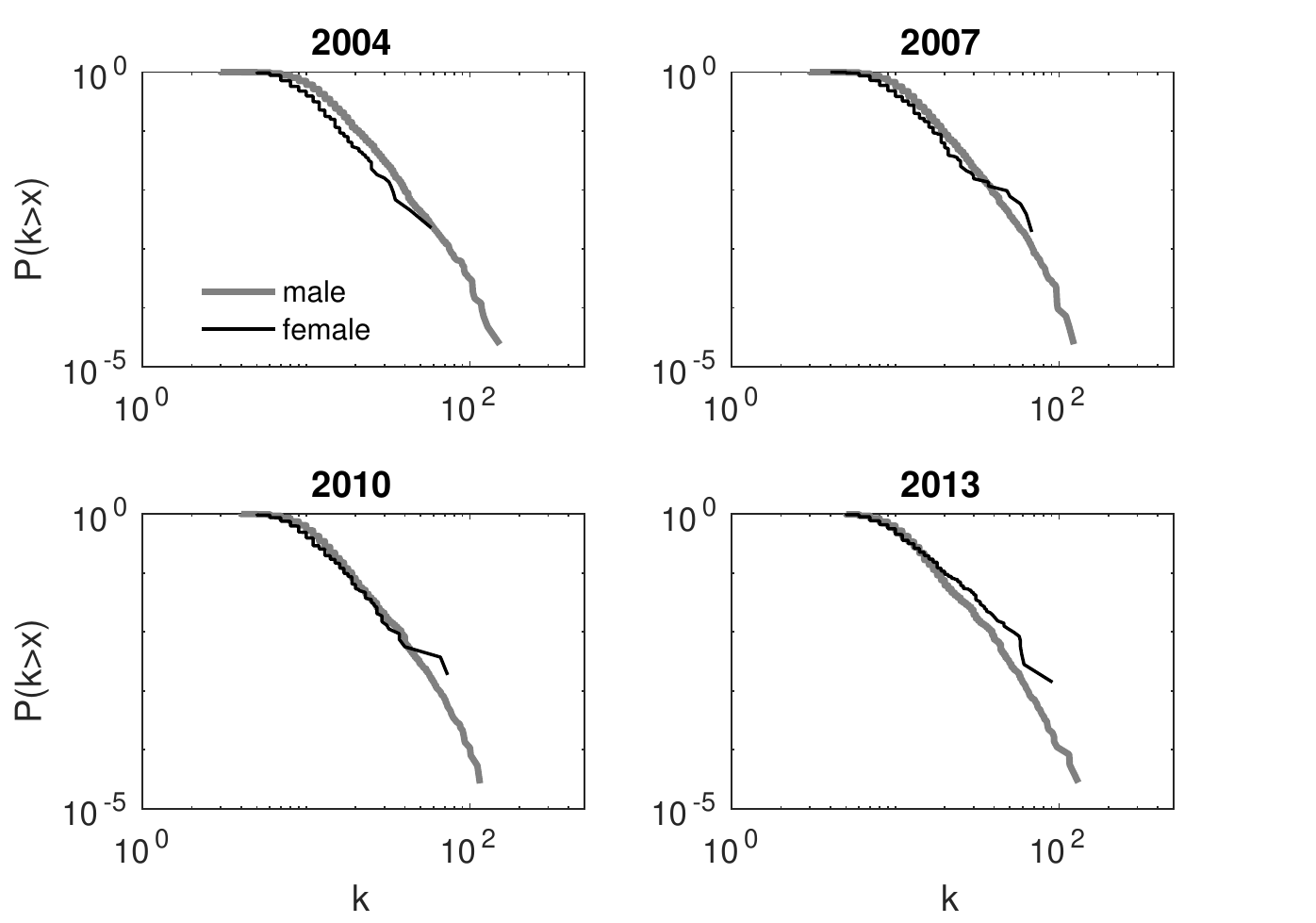}
  \end{center}
  \caption{Degree distributions for male and female board members}{\itshape The figures show the degree distributions on a log-log scale. The distributions for men show some power-law behavior, yet the cut-off is rather high and the tail is somewhat dampened. Although the results for female board members are noisy, a catch-up process in terms of connectvity is visible. }\label{fig:alpha4}
\end{figure}

The increase in the number of mandates naturally leads to an increase in connectivity also in the personal network of board members. To illustrate this we plot the degree distributions of male and female board members in figure \ref{fig:alpha4}. All these distributions are close to power-law behavior, however, one can also clearly see that the underlying networks are not prototypical scale-free graphs. The maximum of connections is clearly limited, likely due to best practices and corporate governance guidelines about multiple mandates. Also, for board members with multiple mandates the number of connections increases step-wise with each board to which they are appointed. This leads to a relatively high cut-off for the power-law and also some distortions at the intermediate range of connectivity.

It is however obvious that female board members catch up to their male counterparts in terms of connectivity. While the maximum for the degree is still significantly lower, the relative number of highly connected female board members in 2013 is even higher for women than for men. It is worth noting that the visible kinks in the distributions are caused by a select few women who stand out from the still relatively small group of female board members.

\begin{figure}[tb]
\footnotesize
\begin{center}
    \includegraphics[width=0.65\textwidth, trim = 40 230 50 200, clip=true]{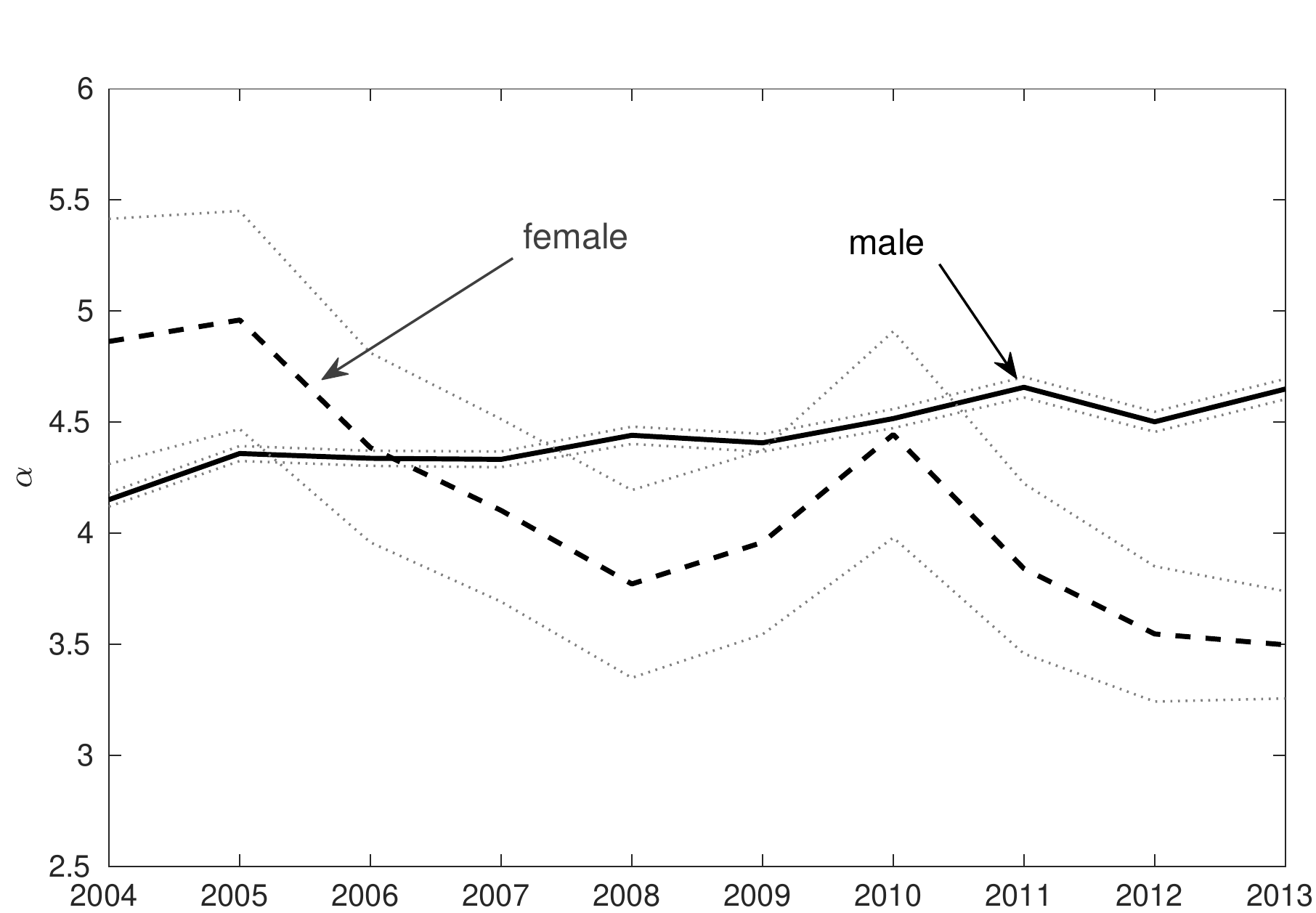}
  \end{center}
  \caption{Estimated alpha parameter for degree distribution of male and female board members over time. }{\itshape The figure compares the estimated $\alpha$ assuming a power-law degree distribution. Dotted lines show the standard error. The results quantitatively confirm that connectivity for female board members has catched up and in fact overtaken that of men. The process seems to have temporarilly reversed in 2009 and 2010.}\label{fig:alpha}
\end{figure}

In order to quantify the changes in these distributions over time we have estimated the tail exponents. The alpha for the fitted power-law distributions are reported in figure \ref{fig:alpha}. For male board members the already high alpha is slightly increasing over time, mirroring the trend of slightly smaller boards and slightly fewer multiple mandates in general. For female board members we see a trend into the opposite direction, their degree distribution becomes more heavy-tailed over time, surpassing those of their male colleagues. 

It is worth noting that such differences in the degree distributions do not only relate to the centrality and visibility of men and women, but that it is also likely an indicator for homophilic behavior.  This has been shown analytically by \cite{eun} in extensions of the scale-free network model of \cite{Barabasi99}. These models produce networks with a tail exponent of 3 in their baseline configuration. However, when different groups are introduced and when these groups are assigned different levels of homophily, tail exponents different from 3 will emerge as a result of this homophilic behavior. A finding like in the first years of our data set, with a higher alpha for the (minority) of women compared to men could be obtained in settings where women have a preference to connect to women while the majority (of men) does not have any preference. It is therefore important to analyze if women (or the boards in general) behave in fact homophilic, since this is likely to affect the dynamics of the network in terms of new board member appointments.

\section{Network effects}\label{sec:nets}
\subsection{Homophily}\label{sec:hom}

As a next step we will investigate if male and female board members are connected to each other in an unbiased way or if the ego networks (the neighborhoods) of men and women differ. In other words, we analyze if homophily exists.

\begin{figure}[htb]
\footnotesize
\begin{center}
    \includegraphics[width=0.75\textwidth, trim = 50 205 50 205, clip=true]{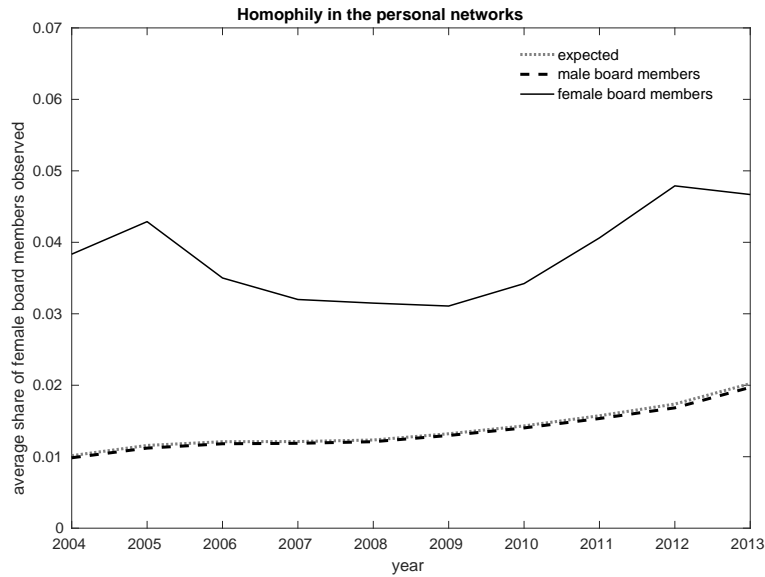}
  \end{center}
  \caption{Homphily in personal networks}{\itshape The figure shown a comparison of the share of female board members in the ego networks of men and women compared with the overall share among all board members. It reveales that female board members connectivity is biased towards being connected to other female board members.}\label{fig:homp}
\end{figure}

We will start by having a look at the personal networks of all board members. We know that the percentage of female board members is close to 1 percent in 2004 and rises to around 2 percent in 2013. This share serves as our benchmark. If ties between men and women were unbiased we expect that the average share of women in all the individual ego networks should be very close to the global share and that it should be identical for men and women. Figure \ref{fig:homp} shows these averages and reveals that female board members are actually connected to more other female board members than what their overall share would suggest. The ratio between the expected and actual share is slightly narrowing over time. For men these shares are almost identical. Their observed share of female board members is only very narrowly below the expected value, a finding which is of course related to the relative group sizes of men and women.\footnote{Different from the predictions of before mentioned analytical models of social networks, there is obviously no reversal of homophily over time in the observed network but only a weakening. This is not completely unexpected because the nodes in our network cannot re-evalute their ties in each year, which leads to persistence in homophily. Also, the empirical network is not a perfect scale-free graph.}

\begin{figure}[tb]
\footnotesize
\begin{center}
    \includegraphics[width=0.75\textwidth, trim = 50 230 60 205, clip=true]{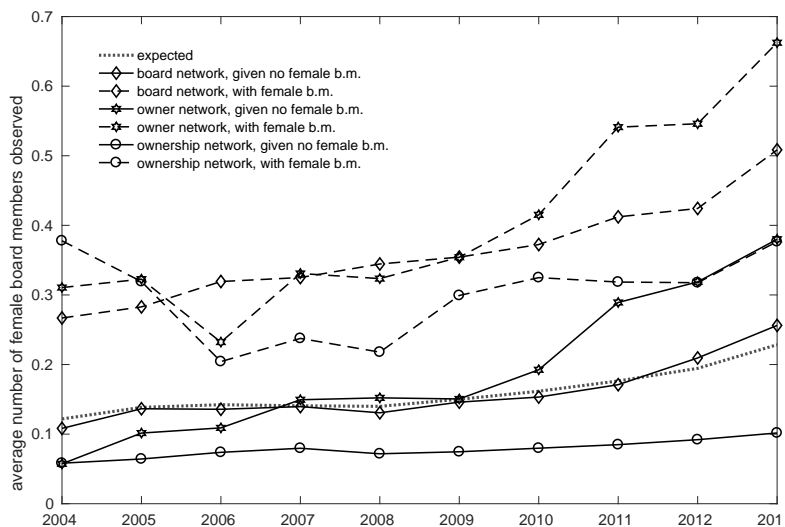}
  \end{center}
  \caption{Homphily in the board network}{\itshape The plot shows the average number of female board members in the ego networks of firms compared with the overall expected value. As networks we consider the board network, as well as the network towards owners and towards owned firms. We condition each case on whether at least one female board member is present on the board of the ego-firm. We find that homophily with repect to the presence of female board members is present also on the level of firms.}\label{fig:homf}
\end{figure}

Personal ties in these networks are obviously mostly not made as an individual decision, they come as a result of membership on the board of a corporation. It therefore makes sense to search for homophily also on the level of firms. We can consider three types of network relationships and their firms' ego networks. First we look at the network between firms based on links in the board network. Here we look at the average number of female board members that serve on the boards of firms in the ego networks. We can compare this figure with the overall expected number of female board members (e.g., in 2004 one would expect 0.12 female board members). Further we can condition this on whether a firm actually has at least one woman on its own board.

We can ask the same question for firms that are linked in the ownership network by shareholding of at least 2 percent\footnote{The threshold of 2\% is required to filter out diversified investments of institutional investors and other
shareholders without significant influence on the owned firm.}. Here we can distinguish between ego networks that connect a firm to its owners and those connections that show the ownership over other firms.\footnote{Since there are firms that are not connected this limits the sample to around 2200 firms for the analysis of the board network and to around 1300 (750) in the owner (ownership) network (the exact numbers vary from year to year).}

Figure \ref{fig:homf} shows that homophily in fact also exists on the level of firms. For all three ego networks the average number of female board members is higher among those alters where the ego has a female board member on its board. For the ownership network these effects seem to depend to some extend on the hierarchy. There are obviously cases where we see female board members on the boards of firms that are owned by somebody else. This does however not necessarily mean that the parent company will also have a female board member. On the other way around, if already the parent company has a female board member, chances are that firms in its ownership ego-network also have a female board member.\footnote{Since the appearance of female board members is close to a binary event we chose in figure \ref{fig:homf} to report the average number of female board members  instead of average percentages (which would relate to varying board sizes). This implies that the  plotted deviation from the expected number of female board members is not symmetric.}

\subsection{Centrality}\label{sec:cent}

Our findings on the degree distributions and the number of mandates have hinted into the direction that -- although in the minority -- some women might have moved into rather central positions in the personal network in the last years of our sample period. In the following we will have a look at this development over time and we will investigate if the distributions of centrality for men and women differ.

\begin{figure}
\footnotesize
\begin{center}
    \includegraphics[width=0.9\textwidth, trim = 40 310 30 280, clip=true]{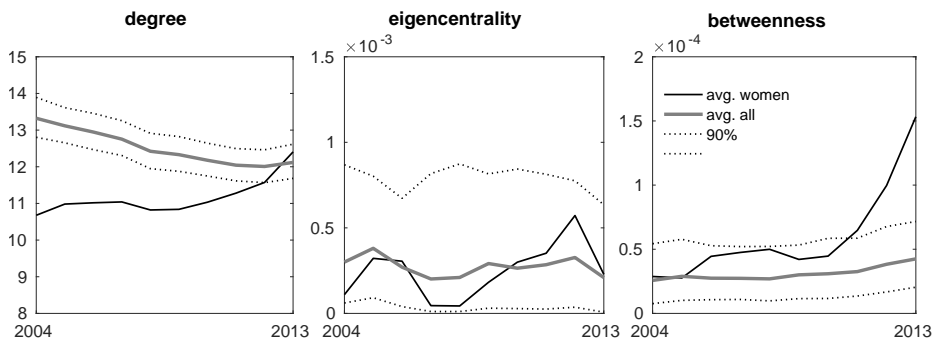}
  \end{center}
  \caption{Average centrality of female board members over time}{\itshape The panels show the average centrality of female board members compared to the overall average in terms of the degree centrality (left panel), eigencentrality (middle panel) and betweenness centrality (right panel). Dotted lines show the 90\% confidence interval for the centrality of a group with identical size drawn from the overall population.}\label{fig:cent_all}
\end{figure}

We start by comparing the average centrality of women over time with the average for all board members. The left panel of figure \ref{fig:cent_all} shows the average degree centrality of female board members. Women do in fact catch up in terms of degree centrality, yet do not pass the overall average by a significant amount. This effect on the average degree can be caused by two effect: women joining firms with larger boards and women having more mandates on average. In order to see whether women have obtained more central positions in general a look at the eigen- and betweenness centrality is necessary (middle and right panel). Due to the high skewness of the first measure we can only report a slightly positive trend in this case, but no significant difference from the overall sample. However, the betweenness measure, which captures the ability of board members to connect otherwise dis-joint parts of the network, shows that those few women that we observe do in fact become noticeably central with respect to betweenness.

\begin{figure}[p]
\footnotesize
\begin{center}
    \includegraphics[width=0.9\textwidth, trim = 60 165 40 150, clip=true]{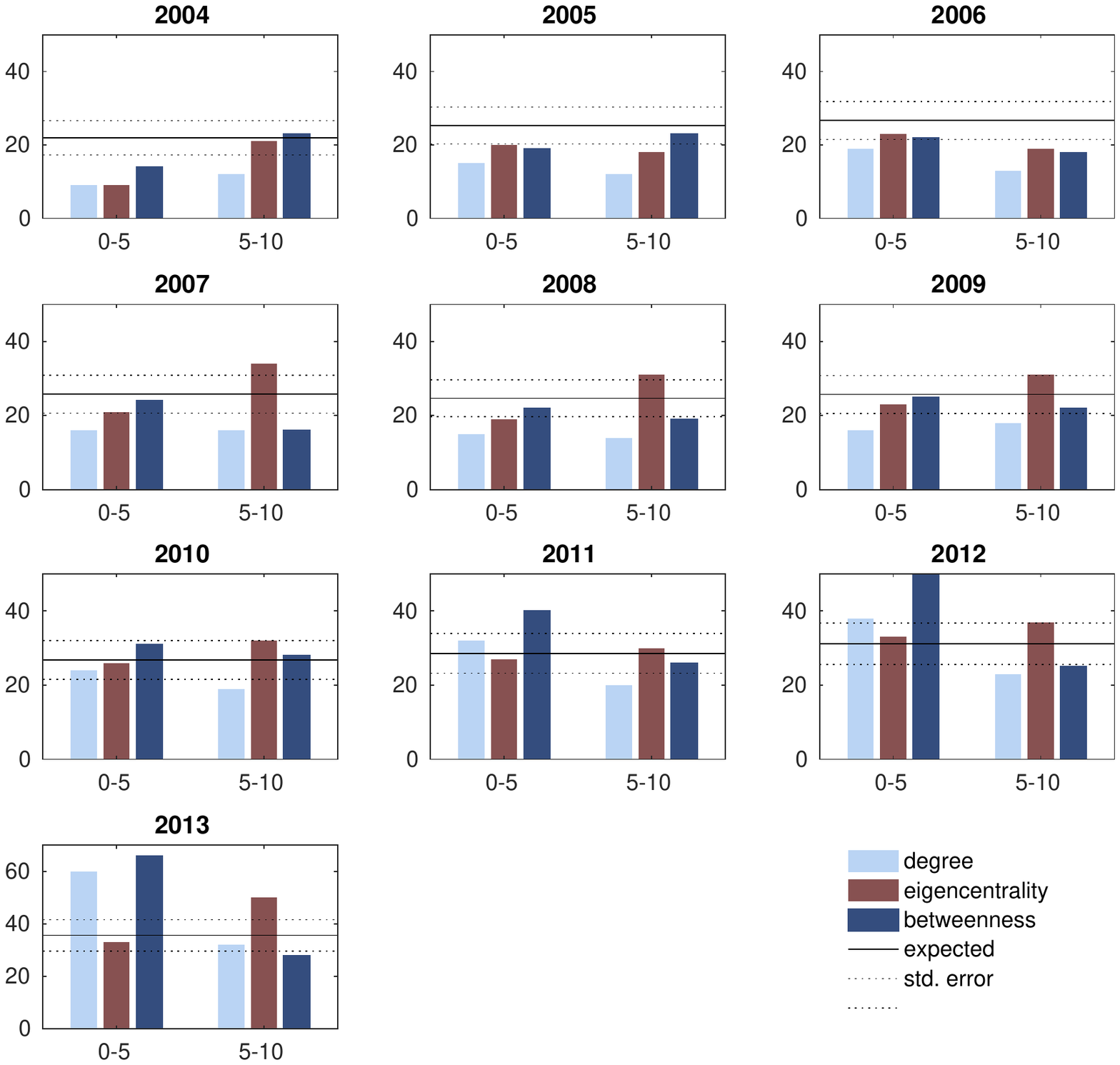}
  \end{center}
  \caption{Representation of women among most central board members}{\itshape The plots show a comparison of the number of female board members who are present in the  top 10\%  of the distributions of degree, eigenvector centrality and betweennes centrality. For each year we have also plotted their expected number and the standard error. While women were underrepresented at least in the top 5\% in 2004, they have caught up over the years, mostly in terms of representation by degree and betweenness centrality.}\label{fig:centr}
\end{figure}

Since distributions of centrality are by nature highly skewed a comparison of averages can be misleading. For the evaluation of centrality in networks it is therefore useful to also investigate the  representation of a group among the most central nodes (conditional on their overall share). 
Such a measure can also serve as a proxy for the general visibility of a group, which depends dis-proportionally on nodes in the top of the distribution, often exacerbated, for example, by recommender systems in social networks and algorithmic selection of media reports \citep{recom,karimirank}.

To analyze the visibility of board members we order them by centrality score and count how many female board members are present in the first two 5 percent intervals of these distributions.\footnote{Due to the heavy tails of these distributions all board members that are outside the top 10 percent are basically equally unimportant. An even more selective definition of the ``top", e.g. statistics on the top 1-2 percent, would be desirable, but would not result in meaningful statistics because of the small number of women in the sample.} We can then compare the actual number of top female board members with their expected numbers, assuming that their centrality should be distributed like that of the entire population. The results are shown in figure \ref{fig:centr}. Due to the low overall number of female board members the results are admittedly noisy, nevertheless, by comparing the results for all the years some trend is visible. When we judge according to degree centrality or eigencentrality   female board members are mostly underrepresented in the top from 2004--2006. When we compare this with the years 2011--2013 we observe that the number of women among the most central players has increased significantly. The results are more pronounced in terms of degree centrality and betweenness centrality than for eigencentrality.

These results suggest that over time many women have moved from positions at smaller firms (with below average board size) to the boards of larger firms. The disparity between the eigencentrality and betweenness values suggest that some very central female board members bridge boards that are otherwise not connected, but that not all of these connections are in the most central part of the network.

\section{Determinants of the number of female board members}\label{sec:numfem}

\begin{table}
\footnotesize
\setlength{\tabcolsep}{2pt}
\begin{tabular}{c | c c c c c c c c c c}
	\hline
	\hline
	year     &  2004 &   2005 &  2006 &    2007 &   2008 &   2009 &   2010 &   2011 &   2012 &   2013 \\ 
	\hline
N     &  3767 &   3849 &  3943 &    3887 &   3767 &   3672 &   3595 &   3543 &   3532 &   3545 \\  
$R^2$    & 0.8941 &  0.8828  &  0.8717  &  0.8747  &  0.8755  & 0.8668  &  0.8573  &  0.8442  &  0.8304  &  0.8011  \\  
$R^2_{f=1}$  &  0.0230 &   0.0338 &  0.0274 &    0.0340 &   0.0417 &   0.0379 &   0.0486 &   0.0593 &   0.0869 &   0.1322 \\  
disp  &  1.0601  &  1.0708  &  1.0563 &  1.0487  & 1.0421  &  1.0176  &  1.0274  &  1.0463  &  1.0532  &  1.0375  \\ 
\hline
\\ 
const &  -2.3855 &   -2.2845 &  -2.2153 &    -2.2699 &   -2.2150 &   -2.1334 &   -2.1197 &   -2.0321 &   -2.0054 &   -1.8839 \\  
      & \tiny{(-21.2843)} & \tiny{(-21.8646)} & \tiny{(-21.7579)} & \tiny{(-22.0459)} & \tiny{(-21.1004)} & \tiny{(-20.6194)} & \tiny{(-20.7458)} & \tiny{(-20.7789)} & \tiny{(-21.3662)} & \tiny{(-21.6469)} \\  
b. size &  -0.0097 &   0.0013 &  0.0116 &    0.0088 &   0.0057 &   0.0036 &   0.0121 &   0.0224 &   0.0257 &   0.0344 \\  
      & \tiny{(-0.7049)} & \tiny{(0.1005)} & \tiny{(0.9690)} & \tiny{(0.7144)} & \tiny{(0.4299)} & \tiny{(0.2648)} & \tiny{(0.9239)} & \tiny{(1.7534)} & \tiny{(2.2049)} & \tiny{(3.3320)} \\  
      \\
degree   &  -0.1015 &   -0.0438 &  -0.1104 &    -0.1102 &   -0.1409 &   -0.0598 &   -0.0539 &   -0.0578 &   0.0223 &   0.0847 \\  
  board    & \tiny{(-1.4903)} & \tiny{(-0.7134)} & \tiny{(-1.7919)} & \tiny{(-1.7110)} & \tiny{(-2.0554)} & \tiny{(-0.9086)} & \tiny{(-0.8438)} & \tiny{(-0.9320)} & \tiny{(0.3820)} & \tiny{(1.5870)} \\  
degree &  -0.0918 &   -0.0165 &  0.0376 &    0.1788 &   0.1953 &   0.1254 &   0.1288 &   0.1056 &   0.1256 &   0.0831 \\  
 owner     & \tiny{(-0.6214)} & \tiny{(-0.1379)} & \tiny{(0.3449)} & \tiny{(1.8574)} & \tiny{(1.9982)} & \tiny{(1.3461)} & \tiny{(1.4675)} & \tiny{(1.2951)} & \tiny{(1.6845)} & \tiny{(1.2486)} \\  
degree &  -0.8420 &   -0.8791 &  -0.7308 &    -0.6985 &   -0.7797 &   -0.7986 &   -0.7455 &   -0.7273 &   -0.7752 &   -0.7734 \\  
 ownsh.     & \tiny{(-5.3768)} & \tiny{(-5.9183)} & \tiny{(-5.5117)} & \tiny{(-5.1929)} & \tiny{(-5.7086)} & \tiny{(-6.3669)} & \tiny{(-6.2143)} & \tiny{(-6.3197)} & \tiny{(-6.9469)} & \tiny{(-7.3946)} \\  
 \\
$<n>$ fem  &  1.1678 &   1.0570 &  1.2839 &    1.1439 &   1.3094 &   1.0265 &   1.0376 &   1.1667 &   1.0755 &   0.9887 \\  
  board net    & \tiny{(5.6676)} & \tiny{(6.1437)} & \tiny{(8.0031)} & \tiny{(7.0344)} & \tiny{(8.0742)} & \tiny{(6.9430)} & \tiny{(7.1847)} & \tiny{(8.5191)} & \tiny{(8.3224)} & \tiny{(8.9032)} \\  
$<n>$ fem &  1.6201 &   1.2621 &  1.0485 &    1.1264 &   0.9493 &   0.9378 &   0.7827 &   0.6668 &   0.7146 &   0.7119 \\  
owner net      & \tiny{(4.4267)} & \tiny{(3.7377)} & \tiny{(3.2129)} & \tiny{(4.0486)} & \tiny{(3.4977)} & \tiny{(3.8465)} & \tiny{(3.7464)} & \tiny{(3.6636)} & \tiny{(4.0270)} & \tiny{(4.7518)} \\  
$<n>$ fem &  1.2501 &   1.3837 &  0.4971 &    0.7906 &   0.9325 &   1.1873 &   1.3334 &   1.1551 &   0.9645 &   1.0685 \\  
owsh net      & \tiny{(3.2547)} & \tiny{(3.9221)} & \tiny{(1.1788)} & \tiny{(2.4181)} & \tiny{(2.5928)} & \tiny{(4.5360)} & \tiny{(5.9166)} & \tiny{(5.5440)} & \tiny{(4.9935)} & \tiny{(6.6107)} \\ 
\\ 
sector  &  0.7011 &   0.7458 &  0.7953 &    0.8791 &   0.6437 &   0.7237 &   0.7294 &   0.7878 &   0.6910 &   0.5473 \\  
 Foods     & \tiny{(3.1538)} & \tiny{(3.7191)} & \tiny{(4.0907)} & \tiny{(4.5886)} & \tiny{(3.0453)} & \tiny{(3.6395)} & \tiny{(3.7678)} & \tiny{(4.3644)} & \tiny{(3.9522)} & \tiny{(3.4288)} \\  
sector &  0.2063 &   -0.0009 &  -0.0233 &    0.1376 &   0.1716 &   0.2888 &   0.2921 &   0.3257 &   0.3865 &   0.2920 \\  
Chem.      & \tiny{(0.8293)} & \tiny{(-0.0037)} & \tiny{(-0.0972)} & \tiny{(0.6221)} & \tiny{(0.7874)} & \tiny{(1.4337)} & \tiny{(1.5255)} & \tiny{(1.8168)} & \tiny{(2.3325)} & \tiny{(1.8991)} \\  
sector &  -0.0388 &   -0.1111 &  -0.1600 &    -0.1440 &   -0.0521 &   -0.1024 &   -0.1534 &   -0.3901 &   -0.3995 &   -0.5925 \\  
  Machinery  & \tiny{(-0.1525)} & \tiny{(-0.4517)} & \tiny{(-0.6678)} & \tiny{(-0.6013)} & \tiny{(-0.2261)} & \tiny{(-0.4546)} & \tiny{(-0.6698)} & \tiny{(-1.6422)} & \tiny{(-1.7686)} & \tiny{(-2.6542)} \\  
sector &  -0.1618 &   -0.3243 &  -0.3977 &    -0.3602 &   -0.3158 &   -0.2635 &   -0.2360 &   -0.2920 &   -0.2038 &   -0.3219 \\  
El. Appl.   & \tiny{(-0.6646)} & \tiny{(-1.3491)} & \tiny{(-1.6613)} & \tiny{(-1.5046)} & \tiny{(-1.3444)} & \tiny{(-1.1929)} & \tiny{(-1.0895)} & \tiny{(-1.3817)} & \tiny{(-1.0486)} & \tiny{(-1.7752)} \\  
sector &  0.7807 &   0.6893 &  0.6187 &    0.5942 &   0.4906 &   0.4538 &   0.5022 &   0.5035 &   0.4880 &   0.3056 \\  
IT    & \tiny{(4.7911)} & \tiny{(4.6183)} & \tiny{(4.1897)} & \tiny{(3.9465)} & \tiny{(3.0894)} & \tiny{(2.9366)} & \tiny{(3.3786)} & \tiny{(3.6147)} & \tiny{(3.6376)} & \tiny{(2.4205)} \\  
sector &  -0.0410 &   0.1289 &  0.1157 &    -0.0125 &   -0.0154 &   -0.1265 &   -0.1117 &   -0.2226 &   -0.1878 &   -0.0490 \\  
WS Trade     & \tiny{(-0.1981)} & \tiny{(0.7099)} & \tiny{(0.6488)} & \tiny{(-0.0660)} & \tiny{(-0.0820)} & \tiny{(-0.6609)} & \tiny{(-0.5989)} & \tiny{(-1.1940)} & \tiny{(-1.0634)} & \tiny{(-0.3241)} \\  
sector &  0.8778 &   0.8489 &  0.8269 &    0.8407 &   0.7335 &   0.6245 &   0.6961 &   0.6787 &   0.6931 &   0.6049 \\  
 R Trade     & \tiny{(5.9324)} & \tiny{(6.1659)} & \tiny{(6.1412)} & \tiny{(6.1531)} & \tiny{(5.2021)} & \tiny{(4.3927)} & \tiny{(5.0443)} & \tiny{(5.1158)} & \tiny{(5.5334)} & \tiny{(5.2330)} \\  
sector &  1.0638 &   1.0080 &  0.9580 &    0.9379 &   0.9424 &   0.9070 &   0.9683 &   0.8985 &   0.8635 &   0.7593 \\  
 Services     & \tiny{(7.2536)} & \tiny{(7.3918)} & \tiny{(7.2551)} & \tiny{(7.0310)} & \tiny{(7.0048)} & \tiny{(6.9315)} & \tiny{(7.5791)} & \tiny{(7.2908)} & \tiny{(7.3024)} & \tiny{(6.9961)} \\  
	
\end{tabular} 
\caption{Determinants of the number of female board members in a firm.}{\itshape Results of the Poisson-regression, t-statistics in parenthesis. As the main infuences we test for connectivity in the board and ownership network \emph{(degree)}, as well as the average number of female board members in the ego-networks, \emph{$<n>$ fem}. Further, 8 sector fixed effects are included to account for industry-specific differences.}\label{tab:fem}
\end{table}

In the following we want to investigate in how far aspects of the network influence the prospects of female board members. Before we go into details it is necessary to point out that there are some other relevant factors that relate to the number of female board members. The first one is the the sector classification of firms. Female board members are more frequently found at firms that provide services or trade than in the more traditional sectors, like mining or construction, see table \ref{tab:secstats} in the appendix. The second factor are differences between family- or founder- versus professionally-managed firms with respect to management \citep[see, e.g.][]{villa}. Although it seems unlikely that this effect is gender-specific we note that in the following we cannot control for this second factor since it was not feasible to manually generate the relevant information for such a large data set. 

Our previous analysis of the network structure, in particular homophily, has revealed that assortativity with respect to gender is related to the number of women on corporate boards. 
Hence, to test the influence of homophily on the composition of the boards we estimate the number of female board members in each firm by coding the connectivity and the exposure to other female board members in the board and ownership networks and by controlling for fixed effects approximated by the sector classification.

Table \ref{tab:fem} reports the results of a Poisson regression that tests for the network effects in the following way. First, the size of the board of each firm has to be considered, since the chance of having a female board member should increase with the number of seats. The variable \emph{b. size} therefore represents the difference in the size of a firm's board from the mean. We test for the influence of connectivity in the board and ownership network by including three variables for the degrees. For the ownership network we consider the in- and out-degree separately (as a logged variable, i.e., $log(1+k_i)$). We further test for the influence of homophily in the networks by accounting for the average number of female board members that is observed at connected firms in either the board or ownership network (for the latter in- and outgoing influences are estimated separately). Further we use 8 dummies, covering the most populated sectors.\footnote{Since in many sectors only very few woman are employed we are limited to the use of about 8 sector dummies for technical reasons. These 8 dummies however cover 61 percent of all firms and we have verified that the exact selection of dummy variables does not influence our findings.}

When we look at the results, the first point to take notice of is that the board size variable starts out with a negative sign and turns to be significantly positive in the later years. As previously discussed, the reason is that many female board members have initially been hired at smaller firms with smaller boards, it is only towards the end of our sample period that also large corporations have hired some female board members \citep[see also][for similar findings]{kubo}. The variable for the degree in the board network (which is mostly not significant) shows a similar behavior. Since large corporations (with very few female board members) are the most networked ones, this finding can be expected. For the ownership network we again find evidence of some hierarchical influences. Firms that have ownership in many other firms have significantly fewer female board members. The influence of the in-degree (connected owners) is tendencially reversed, although in this analysis it is mostly insignificant.

The most important part are of course the variables that describe the average number of female board members in the three respective ego networks. They are all positive and highly significant (with one exception for the ownership network in 2006). This means that the homophily in the network has a measurable impact on the board composition: the number of female board members of a firm is connected to the fact whether female board members are observed at connected peers. Hence, we find evidence for an influence of the corporate networks on the composition of the board with respect to gender.\footnote{To investigate the seemingly good fit of the model we have calculated an additional measure labeled $R^2_{f=1}$ which give the ratio of correct predictions of firms that have exactly one female board member. It shows that these cases are more difficult to predict than having no female board members at all. In particular our model predicts on average 0.1996 female board members for firm with no female board members and 0.3199 (0.4519) for firms with 1(2) female board members.}

\section{Performance}\label{sec:perf}

Previous studies on the relationship of board member gender and financial performance have come to mixed results. As mentioned earlier, most studies assume this to be an indirect relationship \citep{galbreath,kirsch18}. Given the large sample size a simple comparison of corporations with and without female board members should nevertheless give a good indication about differences in financial performance, even if these differences are likely to coincide with differences in governance.

Since we deal with firms from very different sectors we analyze this by looking at the return on assets (ROA) as the most general criterion of performance. Since firms from different sectors will typically have different balance sheet structures, we normalize this ROA by the sector averages in each year. We can then compare the firms that have boards with only men to those where at least one women serves on the board.

\begin{figure}[tb]
\footnotesize
\begin{center}
    \includegraphics[width=0.75\textwidth, trim = 0 240 130 250, clip=true]{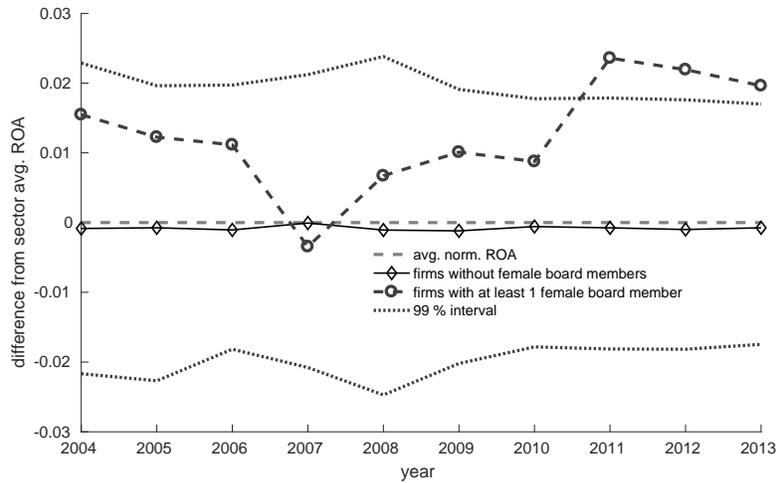}
  \end{center}
  \caption{Firm profitabilty}{\itshape The figure compares the ROA of firms with at least 1 female board member with that of firms without any female board members. The ROAs have been normalized by the annual sector averages. Firms with female baord members tend to have a higher ROA than those without. The dotted lines shows the 99\% confidence interval for the ROA of a group with the size like that of the firms with female board members.}\label{fig:prof}
\end{figure}

For this analysis we use a sub-sample of 1,357 firms for which data is available for all 10 years. 1,187 of these firms never had a female board member.  We will focus on the 32 firms that have at least one female board member on average over this time period and compare their performance to the sample average (we obtain qualitatively very similar results as long as the average number of female board members is $>0.7$).

Our results for this analysis are shown in figure \ref{fig:prof}. We can see that the ROA for the majority of firms with entirely male boards is very close to the overall average. For the firms with at least one female board member we have numerically determined the 99-percent interval, i.e., the expected range for the average ROA of a randomly select sub-sample of firms. Interestingly we find that firms with female board members do perform better most of the time. This result is however only significant in the last three years. 

Given that the majority of studies on board composition struggle to establish any relationship to performance this finding is in fact surprising. It should however not be overstated. As already mentioned, it is likely that this finding is related to the quality of governance in general. Firms that have good governance are more profitable and those firms might also be better in facilitating the hiring of female board members. In this respect our results are similar to those of \cite{farrell} who also found women to serve for better performing firms. The results also support \cite{camp} who found that increasing the number of female board members is not an issue for firm performance. Given that the overall number of female board members is still low, it seems plausible that effects related to inexperienced board members, as described by \cite{ahdit}, should not be an issue in the Japanese case.

\section{Appointment of female board members}\label{sec:hire}

After looking at the networks of existing female board members and firm performance we have to look at the dynamics of the appointment process. The aim is to find out if network effects also explain this process and could therefore be useful for policy advice on gender diversity (a visualization of appointments in the network is provided in figure \ref{fig:viz} in the appendix).

In order to look at the determinants of female hirings we again limit the analysis to the 1,357 corporations which are present throughout the entire time period and for which data on earnings and shareholder-ship is available. In the time period from 2004--2013 218 new female board members have been hired by 179 corporations. We compare the firms which have hired female board members with those that did not hire any and we analyze in how far network effects and firm characteristic can explain this behavior.

\begin{table}[p]
\footnotesize
\begin{center}
\begin{tabular}{c | c c c c c c c c c c}
	\hline
	\hline
 model        & \multicolumn{2}{c}{only year dummies}   & \multicolumn{2}{c}{incl. sectors} &   \multicolumn{2}{c}{incl. foreign sh.}     \\ 
	\hline
	MF $R^2$ & 0.1293 & & 0.1455 && 0.1470 \\
	LR ratio & 275.9 && 310.5 && 313.6  \\
	log L & -929.0 && -911.8 && -910.2 \\
	obs. / ones & 10820 & 218 & 10820 & 218 & 10820 &218 \\
	vars & 18  && 25 && 26 \\
	\hline
	\\
year dummies & $\bullet $ & & $\bullet $ & & $\bullet $ & \\ 
\\
board size  &  0.0428 &   (2.72) &     0.0488 &    (3.03)  &  0.0466 &    (2.89) \\ 
ROA   &    4.2788 &   (2.62)   &  2.9835 &   (1.76)     &    2.0992   &  (1.20) \\
foreign sh. &  & & & &  1.2164  &    (1.80) \\  
\\
degree board &   0.1931 &  (1.86) &   0.2258  &    (2.15)  &  0.2107 &    (2.00) \\
degree owner &  0.4056  &  (3.87)  &   0.4474 &   (4.19) &   0.3979 &     (3.62)   \\
degree ownsh. &   -0.7653 & (-4.95) & -0.8250 &   (-5.23)  &  -0.7603 &  (-4.69)  \\  
\\
fem. board net &   0.6445 &  (1.88) &   0.6222  & (1.80)  &   0.5767   & (1.66)   \\
fem. owner net &   1.1913 &  (4.10) &  1.2761 &   (4.35)  &  1.2806 &   (4.37) \\
fem. ownsh. net & 1.2522 &  (3.26)  &  0.9763 &   (2.50)  &  0.9909 &   (2.53) \\
existing fem. &  1.4977 &  (7.56)   &  1.3068 &   (6.42)  &  1.3232 &   (6.50) \\
\\
Foods  &  & & 1.2219 & (4.36) &    1.2649 &   (4.49) \\
Chemicals & & &  0.7689 &  (3.24) &  0.7521 &  (3.17) \\
Pharmac. & & &  1.0618 &  (2.61)  &  1.0625  &  (2.63) \\
IT  & & &   0.6379 &  (2.19) &    0.6738  &     (2.30) \\
WHS Trade & & &  0.7148 &   (3.15) &     0.7443  &  (3.27) \\
RT Trade & & &  0.6232 &    (1.83) &     0.6715  &  (1.97) \\
Services & & &  0.9886 &    (3.39) &     1.0547 &    (3.60) \\

\end{tabular}
\end{center}
\caption{Determinants of new female board members.}{\itshape The table shows results of the logistic regression, t-statistics in parentheses. We compare the characteristics of firms with respect to wheater they hire an additional female baord member in the period from 2004--2013. We estimate three models that differ by the inclusion of sector-fixed effects and a variable that represents foreign share holdings. We find that, besides sector-specific differences, the relative position in the ownership network as well as the presence of other female board members at the boards of connected firms significantly influence the appointment of female board members.}\label{tab:newfem}

\end{table}

Table \ref{tab:newfem} shows the results of a logistic regression where the dependent variable signals the hiring of a female board member (based on the changes observed in the annual reports).
We estimate three versions of this model. All of the models contain fixed effects for the different years. They also contain the two firm characteristics that, as we have seen before, might relate to hiring, namely the \emph{board size} and the profitability, the \emph{ROA}, of the company.

All three versions contain data on the number of connections in the board network, the in- and out-degree in the ownership network and the average number of observed female board members in the respective ego-networks,\footnote{Effectively, we are looking at 9 transition from year $t$ to $t+1$. When we observe the appointment of a female board in the annual report published in $t+1$ we relate this to the characteristics of the firm in year $t$ since this is the year in which the appointment was most likely decided upon.} formatted in the same way as described in section \ref{sec:numfem}. Additionally we consider the case that some firms might already have a female board member and hire another one. This is accounted for by the \emph{existing fem} variable.

The left columns show the results for this baseline model. The model shown in the middle columns adds fixed effects based on sector classification.\footnote{Also here, the low number of female board members limits the possible number of sector dummy variables. We have included the most populated sectors with the most hirings. We have verified that the results do not depend on the exact choice of sectors, in particular we have also estimated the model with the exact same set of sector dummies as in table \ref{tab:fem} and found no significant difference in the results. Our choice of sectors differs partly from the model presented in section \ref{sec:numfem} since here we look at differences in the change of female board membership whilst we looked at differences in levels before.} The two rightmost columns also include the effects of the percentage share of foreign shareholder-ship. 

The results show that network effects are significantly related to the appointment of female board members. The results are robust with respect to the use of sector dummies. The seven network variables are all significant, four of them even at the $\alpha < 1$ \% level. The influences are weaker in the board network than in the ownership network. In the latter we observe a similar hierarchical pattern as before, companies which have some control over others (\emph{degree ownsh.}) tend to have fewer female board members. This effect can however be weakened if those controlled companies do have female board members (\emph{fem. ownsh. net}). Companies with an existing female board member are always more likely to appoint additional ones.

We can follow up on our previous observations about profitability by looking at the effect of the \emph{ROA}. While its coefficient is positive in all three models, we also observe that by including the share of foreign shareholding the effect becomes insignificant. One can argue that \emph{foreign sh.} can serve as a proxy for a corporations' openness to modern (`western') corporate governance practices, although it is also slightly correlated to the size of firms in general. However, the results for \emph{ROA} and \emph{foreign sh.} hint into the direction that besides network effects governance is of course related to both, firm performance and the composition of the board.

Since our model estimates the marginal effects on probabilities we can also have a look at the conditional in-sample prediction accuracy. This means to check if the 218 events that our model classifies as the most likely to lead to an appointment actually do so. While the probability for all events is 2.01 percent, the top 218 predictions turn out to be correct in 24.31 percent of the cases. We can also ask the same question on the level of firms and take as given that the appointments have taken place at 179 firms and check if the 179 firms that our model predicts to be the most likely ones associated with an appointment do hire at least one female board member. What we find is that while the unconditional probability to hire a female board members among all firms during the 10-year time span is 13.19 percent, the 179 firm predicted by the model are correct 18.99 percent of the time. (Since firms that do not hire appear in the sample each year and those that do only appear when they hire, these two figures differ.)

\section{Conclusions}

Our analysis has shown that even if the number of female board members in Japan is still low, they have started to change their position in the networks of executives. Relatively many female board members have multiple mandates, which made them become more central. 

Network effects play an important role in explaining the gender ratio of corporate boards and they are also an important aspect of new appointments. New female board members are often appointed when connected firms also have at least one female board member. In the ownership networks this effect is hierarchical. The relationship of female hirings with performance is tendencially positive.

What remains to be investigated is if the network effects among (minimally) diversified boards will continue to work in favor of an improvement in  gender equality. From a theoretical point of view one should probably avoid too homophilic structures, since these could shield conservative sectors from change and would thus limit the relative centrality and thus influence of female executives. Policy measures to increase the number of women on corporate boards should therefore not only target their overall number but also aim for their appointments to take part in all parts of the economy.

\section*{Acknowledgments}
We thank Chieko Takahashi for help in compiling the board network data. We thank Fariba Karimi for discussing a first draft. MR is grateful for partial funding of this work by the Japan Society for the Promotion of Science (JSPS KAKENHI, JP20K01751).

\section*{References}
\bibliographystyle{elsarticle-harv} 
\bibliography{network2}

\begin{thebibliography}{46}
\expandafter\ifx\csname natexlab\endcsname\relax\def\natexlab#1{#1}\fi
\expandafter\ifx\csname url\endcsname\relax
  \def\url#1{\texttt{#1}}\fi
\expandafter\ifx\csname urlprefix\endcsname\relax\def\urlprefix{URL }\fi

\bibitem[{Ahern and Dittmar(2012)}]{ahdit}
Ahern, K.~R., Dittmar, A.~K., 2012. The changing of the boards: the impact on
  firm valuation of mandated female board representation. The Quarterly Journal
  of Economics 127~(1), 137--197.

\bibitem[{Barab\'asi and Albert(1999)}]{Barabasi99}
Barab\'asi, A.-L., Albert, R., 1999. Emergence of scaling in random networks.
  Science 286, 509--512.

\bibitem[{Bilimoria(2006)}]{bilimoria}
Bilimoria, D., 2006. The relationship between women corporate directors and
  women corporate officers. Journal of Managerial Issues 18, 47--61.

\bibitem[{Borgatti and Foster(2003)}]{Borgatti03}
Borgatti, S.~P., Foster, P.~C., 2003. The network paradigm in organizational
  research: A review and typology. Journal of Management 29~(6), 991--1013.

\bibitem[{Bramoullé et~al.(2012)Bramoullé, Currarini, Jackson, Pin, and
  Rogers}]{bram}
Bramoullé, Y., Currarini, S., Jackson, M.~O., Pin, P., Rogers, B.~W., 2012.
  Homophily and long-run integration in social networks. Journal of Economic
  Theory 147~(5), 1754--1786.

\bibitem[{Branson(2006)}]{branson}
Branson, D., 2006. No Seat at the Table: How Corporate Governance keeps Women
  out of {America’s} boardrooms. New York University Press, New York.

\bibitem[{Brinkhuis and Scholtens(2018)}]{brink}
Brinkhuis, E., Scholtens, B., 2018. Investor response to appointment of female
  {CEOs} and {CFOs}. The Leadership Quarterly 29~(3), 423--441.

\bibitem[{Buchanan and Deakin(2009)}]{jpn_gov}
Buchanan, J., Deakin, S., 2009. In the shadow of corporate governance reform:
  {Change} and continuity in managerial practice at listed companies in
  {Japan}. In: Whittaker, D.~H., Deakin, S. (Eds.), Corperate Governance and
  managerial Reform in {Japan}. Oxford UP, pp. 28--69.

\bibitem[{Campbell and Minguez-Vera(2008)}]{camp}
Campbell, K., Minguez-Vera, A., 2008. Gender diversity in the boardroom and
  firm financial performance. Journal of Business Ethics 83, 435--451.

\bibitem[{Cao et~al.(2021)Cao, Li, Li, Zeng, and Zhou}]{board_diverse}
Cao, C., Li, X., Li, X., Zeng, C., Zhou, X., 2021. Diversity and inclusion:
  Evidence from corporate inventors. Journal of Empirical Finance 64, 295--316.

\bibitem[{Chattopadhyay et~al.(2004)Chattopadhyay, Tluchowska, and
  George}]{ingroup}
Chattopadhyay, P., Tluchowska, M., George, E., 2004. Identifying the ingroup: A
  closer look at the influence of demographic dissimilarity on employee social
  identity. Academy of Management 29~(2), 180--202.

\bibitem[{Chen et~al.(2018)Chen, Leung, and Evans}]{fem_rep}
Chen, J., Leung, W.~S., Evans, K.~P., 2018. Female board representation,
  corporate innovation and firm performance. Journal of Empirical Finance 48,
  236--254.

\bibitem[{Cimpian et~al.(2020)Cimpian, Kim, and McDermott}]{cimpian}
Cimpian, J.~R., Kim, T.~H., McDermott, Z.~T., 2020. Understanding persistent
  gender gaps in {STEM}. Science 268~(6497), 1317--1319.

\bibitem[{Conyon and He(2017)}]{cony17}
Conyon, M.~J., He, L., 2017. Firm performance and boardroom gender diversity: A
  quantile regression approach. Journal of Business Research 79, 198--211.

\bibitem[{Dalton and Dalton(2011)}]{dalton}
Dalton, D.~R., Dalton, C.~M., 2011. Integration of micro and macro studies in
  governance research: {CEO} duality, board composition, and financial
  performance. Journal of Management 37~(2), 404--411.

\bibitem[{Devos et~al.(2009)Devos, Prevost, and Puthenpurackal}]{devos}
Devos, E., Prevost, A., Puthenpurackal, J., 2009. Are interlocked directors
  effective monitors? Financial Management 38, 861--887.

\bibitem[{Dezs{\"o} and Ross(2012)}]{dezso}
Dezs{\"o}, C.~L., Ross, D.~G., 2012. Does female representation in top
  management improve firm performance? {A} panel data investigation. Strategic
  Management Journal 33~(9), 1072--1089.

\bibitem[{Faccio et~al.(2016)Faccio, Marchica, and Mura}]{faccio}
Faccio, M., Marchica, M., Mura, R., 2016. {CEO} gender, corporate risk-taking,
  and the efficiency of capital allocation. Journal of Corporate Finance 39,
  193--209.

\bibitem[{Farrell and Hersch(2005)}]{farrell}
Farrell, K., Hersch, P., 2005. Additions to corporate boards: the effect of
  gender. Journal of Corporate Finance 11, 85--106.

\bibitem[{Galbreath(2021)}]{galbreath}
Galbreath, J., 2021. Is board gender diversity linked to financial performance?
  {The} mediating mechanism of {CSR}. Business \& Society 57, 863--889.

\bibitem[{{Gender Equality Bureau Cabinet Office}(2017)}]{genderE}
{Gender Equality Bureau Cabinet Office}, 2017. Status of female officers in
  listed companies (in {Japanese}).
  Http://www.gender.go.jp/policy/mi\\eruka/company/yakuin.html, retrieved
  04/2018.

\bibitem[{Gulati and Gargiulo(1999)}]{gulati}
Gulati, R., Gargiulo, M., 1999. Where do interorganizational networks come
  from? American Journal of Sociology 104~(5), 1439--1493.

\bibitem[{Horton et~al.(2012)Horton, Millo, and Serafeim}]{horton}
Horton, J., Millo, Y., Serafeim, G., 2012. Resources or power? {Implications}
  of social networks on compensation and firm performance. Journal of Business
  Finance \& Accounting 39, 399--426.

\bibitem[{Jadidi et~al.(2017)Jadidi, Karimi, Lietz, and Wagner}]{jadidi}
Jadidi, M., Karimi, F., Lietz, H., Wagner, C., 2017. Gender disparities in
  science? {Dropout}, productivity, collaborations and success of male and
  female computer scientists. Advances in Complex Systems 1750011.

\bibitem[{Karimi et~al.(2018)Karimi, G{\'e}nois, Wagner, Singer, and
  Strohmaier}]{karimirank}
Karimi, F., G{\'e}nois, M., Wagner, C., Singer, P., Strohmaier, M., 2018.
  Homophily influences ranking of minorities in social networks. Scientific
  Reports 8~(11077).

\bibitem[{Kirsch(2018)}]{kirsch18}
Kirsch, A., 2018. The gender composition of corporate boards: A review and
  research agenda. The Leadership Quarterly 29~(2), 346--364.

\bibitem[{Kirsch(2021)}]{kirsch21}
Kirsch, A., 2021. Revolution from above? {Female} directors’ equality-related
  actions in organizations. Business \& Society, 00076503211001843.

\bibitem[{Kolev(2012)}]{gue}
Kolev, G.~I., 2012. Underperformance by female {CEOs}: A more powerful test.
  Economics Letters 117~(2), 436--440.

\bibitem[{Krichene et~al.(2019)Krichene, Fujiwara, Chakraborty, Arata,
  Hiroyasu, and Terai}]{son1}
Krichene, H., Fujiwara, Y., Chakraborty, A., Arata, Y., Hiroyasu, I., Terai,
  M., 2019. The emergence of properties of the {Japanese} production network:
  {How} do listed firms choose their partners? Social Networks 59, 1--9.

\bibitem[{Kubo and Nguyen(2021)}]{kubo}
Kubo, K., Nguyen, T. P.~T., 2021. Female {CEO} on {Japanese} corporate boards
  and firm performance. Journal of The Japanese and International Economies,
  101163.

\bibitem[{Lamb and Roundy(2016)}]{lamb}
Lamb, N.~H., Roundy, P., 2016. The ``ties that bind'' board interlocks
  research: a systematic review. Management Research Review 39~(11),
  1516--1542.

\bibitem[{Lee et~al.(2019)Lee, Karimi, Wagner, Jo, Strohmaier, and
  Galesic}]{eun}
Lee, E., Karimi, F., Wagner, C., Jo, H.-H., Strohmaier, M., Galesic, M., 2019.
  Homophily and minority-group size explain perception biases in social
  networks. Nature Human Behavior 3, 1078--1087.

\bibitem[{Lincoln et~al.(1992)Lincoln, Gerlach, and Takahashi}]{dyad}
Lincoln, J.~R., Gerlach, M.~L., Takahashi, P., 1992. A dyad analysis of
  intercorporate ties. American Sociological Review 57~(5), 561--585.

\bibitem[{Matsa and Miller(2011)}]{matsa}
Matsa, D.~A., Miller, A.~R., 2011. Chipping away at the glass ceiling: Gender
  spillovers in corporate leadership. American Economic Review 101~(3),
  635--39.

\bibitem[{Mcdowell and Smith(1992)}]{mcd}
Mcdowell, J., Smith, J., 1992. The effect of gender‐sorting on propensity to
  coauthor: Implications for academic promotion. Economic Inquiry 30~(1),
  68--82.

\bibitem[{Mizruchi(1996)}]{mizruchi}
Mizruchi, S., 1996. What do interlocks do? {An} analysis, critique, and
  assesment of research on interlocking directorates. Annual review of
  Sociology, 271--298.

\bibitem[{Page(2007)}]{page}
Page, S.~E., 2007. The Difference: How the Power of Diversity Creates Better
  Groups, Firms, Schools, and Societies. Princeton, NJ: Princeton University
  Press.

\bibitem[{Provan et~al.(2007)Provan, Fish, and Sydow}]{Provan07}
Provan, K.~G., Fish, A., Sydow, J., 2007. Interorganizational networks at the
  network level: A review of the empirical literature on whole networks.
  Journal of Management 33~(3), 479--516.

\bibitem[{Raddant and Takahashi(2021)}]{radhiro1}
Raddant, M., Takahashi, H., 2021. Corporate boards, interorganizational ties
  and profitability: {The} case of {Japan}. Empirical Economics, forthcoming.

\bibitem[{Saito(2017)}]{saito}
Saito, Y., 2017. Female board of directors and organisational diversity in
  {Japan}. CEAFJP Discussion Paper Series, 17--05.

\bibitem[{Solimente et~al.(2017)Solimente, Coluccia, and Fontana}]{femita}
Solimente, S., Coluccia, D., Fontana, S., 2017. Gender diversity on corporate
  boards: an empirical investigation of {Italian} listed companies. Palgrave
  Communications 3~(1), 16109.

\bibitem[{Torchia et~al.(2011)Torchia, Calbro, and Huse}]{femnorw}
Torchia, M., Calbro, A., Huse, M., 2011. Women directors on corporate boards:
  From tokenism to critical mass. Journal of Business Ethics 102~(2), 299--317.

\bibitem[{Uzzi(1996)}]{uzzi}
Uzzi, B., 1996. The sources and consequences of embeddedness for the economic
  performance of organizations: The network effect. American Sociological
  Review 61~(4), 674--698.

\bibitem[{Villalonga and Amit(2006)}]{villa}
Villalonga, B., Amit, R., 2006. How do family ownership, control and management
  affect firm value? Journal of Financial Economics 80~(2), 385--417.

\bibitem[{Zhou et~al.(2012)Zhou, Xu, Li, Josang, and Cox}]{recom}
Zhou, X., Xu, Y., Li, Y., Josang, A., Cox, C., 2012. The state-of-the-art in
  personalized recommender systems for social networking. Artificial
  Intelligence Review 37, 119--132.

\bibitem[{Zona et~al.(2018)Zona, Gomez-Mejia, and Withers}]{zona}
Zona, F., Gomez-Mejia, L.~R., Withers, M.~C., 2018. Board interlocks and firm
  performance: towards a combined agency-resource dependence perspective.
  Journal of Management 44~(2), 586--618.

\end{thebibliography}
\newpage
\begin{appendix}

\section{Female board members over time}


\begin{table}[htb!]
\footnotesize
  \begin{center}
  \setlength{\tabcolsep}{2pt}
    \begin{tabular}{l c c c c c c c c }
      \hline
    &   \multicolumn{4}{c}{2004} & \multicolumn{4}{c}{2013} \\
      sector & firms & b. m. & female b.m. & share & firms & b. m. & female b.m. & share \\
      \hline
 Insurance  &	11 &	179 &	6 &	0.0335 &	12 &	147 &	12 &	0.0816 \\
Services  &	325 &	3463 &	88 &	0.0254 &	359 &	3612 &	148 &	0.0410  \\
Air Transportation    &	6 &	82 &	0 &	0 &	6 &	74 &	3 &	0.0405 \\
Electric Power and Gas   &	25 & 	470 &	7 &	0.0149 &	23 &	366 &	14 &	0.0382 \\
Oil and Coal Prod.   &	13 &	172 &	1 &	0.0058 &	13 &	163 &	6	& 0.0368 \\
Retail Trade   &	375 &	4163 &	85 &	0.0204 &	342 &	3617 &	121 &	0.0335 \\
Rubber Products   &	21 &	260 & 	2 &	0.0077 &	19 &	221 &	7	& 0.0317 \\
Foods   &	157 &	1988 &	25 &	0.0126 &	131 &	1563 &	48 &	0.0307 \\
 Fishery, Agric. For. &	11 &	146 &	3 &	0.0205 &	11 &	135 &	4 &	0.0296 \\
Real Estate &	110 &	1172 &	24 &	0.0205 &	111 &	1084 &	30 &	0.0277 \\
Pharmaceutical   &	54 &	651 &	7	 & 0.0108 &	61 &	698 &	18 &	0.0258 \\
Information and  Comm.  &	306 &	3301 &	63 &	0.0191 &	349 &	3517 &	89 &	0.0253 \\
Other Financing B.    &	55 &	648 &	7	& 0.0108 &	32 &	381 &	8 &	0.0210 \\
Chemicals   &	216 &	2707 &	19 &	0.0070 &	215 &	2496 &	52 &	0.0208 \\
Banks    &	96 &	1382 &	2 &	0.0014 &	93 &	1408 &	29 &	0.0206 \\
\textit{2013 average} & & & & & & & & \textit{0.0194} \\
 Textiles App.   &	87 &	987	& 7 &	0.0071 &	56 &	625 &	11 &	0.0176 \\
Securities and Com. F.   &	42 &	506 &	2 &	0.0040 &	40 &	414 &	7	 & 0.0170 \\
Precision Instruments &	51 &	575 &	3	 & 0.0052 &	50 &	560 &	9 &	0.0161 \\
Other Products &	111 &	1345 &	13 &	0.0100 &	109 &	1202 &	19 &	0.0158 \\
Glass and Cer. &	72 &	809 &	4 &	0.0049 &	62 &	661 &	10 &	0.0151 \\
Wholesale Trade   &	375 &	4319 &	31 &	0.0072 &	343 &	3656 &	54 &	0.0148 \\
Pulp and Paper  &	29 &	366 &	1 &	0.0027 &	26 &	303 &	4 &	0.0132 \\
Electric Appl.  &	305 &	3676 &	20 &	0.0054 &	273 &	3066 &	35 &	0.0114 \\
Land Transportation   &	70 &	1038 &	4 &	0.0039 &	65 &	930 &	9 &	0.0097 \\
Warehousing and H. T.   &	43 &	573 &	1 &	0.0017 &	41 &	519 &	5 &	0.0096 \\
Machinery &	249 &	2951 &	18 &	0.0061 &	230 &	2563 &	22 &	0.0086 \\
Metal Products &	100 &	1143 &	3 &	0.0026 &	91 &	976 &	8 &	0.0082 \\
Nonferrous Metals &	39 &	458 &	0 &	0 &	36 &	394 &	3 &	0.0076 \\
Transportation Eq. &	106 &	1545 &	4 &	0.0026 &	100 &	1225 &	9 &	0.0073 \\
Construction &	223 &	2945 &	5 &	0.0017 &	174 &	2075 &	12 &	0.0058 \\
 Marine Transportation &	19 &	249 &	1 &	0.0040 &	16 &	188 &	1 &	0.0053 \\
Iron and Steel  &	56 &	732 &	2 &	0.0027 &	49 &	568 &	3 &	0.0053 \\
Mining  &	9 &	118 &	1 &	0.0085 &	7 &	96 &	0 &	0 \\

    \end{tabular}
  \end{center}
  \caption{Board members statistics by sector}{\itshape We compare the number of female board members in 2004 and 2013. The rows are sorted by the sector-average in 2013. We state the number of firms in each sector together with the number of board members and the share of women.}\label{tab:secstats}
\end{table}

\begin{figure}[!htb]
\footnotesize
\begin{center}
    \includegraphics[width=\textwidth, trim = 200 270 190 320, clip=true]{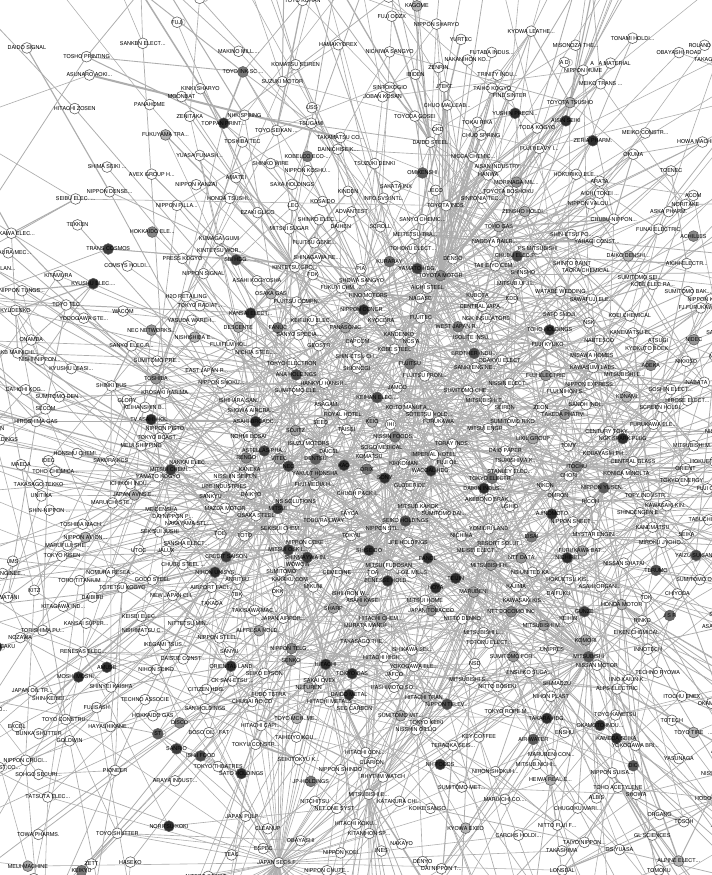}
  \end{center}
  \caption{Female board members in the corporate network}{\itshape  Links resemble connectivity on the board and ownership network within the period 2004--2013. Nodes (firms) that appear white do not have any female board member. Nodes in shades of grey have appointent at least one female board member, the node is the darker the earlier the appointment took place.}\label{fig:viz}
\end{figure}

\end{appendix}

\end{document}